\RequirePackage{lineno} 
\documentclass[aps,prd,preprint,amssymb,groupedaddress]{revtex4-2} %for arXiv
\usepackage{graphicx}% Include figure files
\usepackage{subcaption}
\usepackage{dcolumn}% Align table columns on decimal point
\usepackage{bm}% bold math
\usepackage[usenames]{color}
\usepackage{multirow}
\usepackage{amsmath}
\usepackage{bigstrut}
\def\met{\mbox{\ensuremath{\, \slash\kern-.6emE_{\rm T}}}}

\begin{document}

%\preprint{APS/123-QED}

\title{Limits on 331 vector bosons from LHC proton collision data
}

%\altaffiliation[Also at ]{Instituto de F\'{\i}sica, Universidade Federal do Rio de Janeiro}%Lines break automatically or can be forced with \\
\author{A. A. Nepomuceno}%
\email{andrenepomuceno@id.uff.br}

\affiliation{%
Departamento de Ci\^encias da Natureza\\
Universidade Federal Fluminense \\
Rua Recife s/n,Rio das Ostras, Rio de Janeiro, Brazil
}%

\author{B. Meirose}
\email{Bernhard.Meirose@cern.ch}
\affiliation{
Department of Physics \\
University of Texas at Dallas, USA \\
800 W Campbell Rd, Richardson, TX 75080
}%

\date{\today}% It is always \today, today,

\begin{abstract}
%\noindent 
In this paper, limits are set on bileptons masses and couplings in the context of the 
%minimal and non-minimal versions of 
the 331 Model, as well as generic, non-331 Models predicting bileptons. The following measurable processes are studied: $pp \rightarrow \ell^{+}\ell^{+}\ell^{-}\ell^{-}X$, $pp \rightarrow \ell^{+}\ell^{-} \nu \nu X$ and $pp \rightarrow \ell^{+}\ell^{-} X$.
Experimental limits on singly-charged bileptons masses and couplings within 331 Models are also obtained for the first time. With the results, an over 20 year-old experimental limit on vector bileptons is increased by 60\%. The computed limits are now the most stringent ones for these particles.
\end{abstract}
\pacs{12.60.Cn, 14.70.Pw}
%PACS
%12.60.Cn: Extensions of electroweak gauge sector
%14.70.Pw: Other gauge bosons

\keywords{bileptons} %add keyword for Z´ @419

\maketitle
%\linenumbers
\section{\label{sec:level1}Introduction\protect\\
} 

%\section{Motivations}
\par

%\subsection{331 Models}

The extended elecroweak gauge symmetry group $SU(3)_C \otimes SU(3)_L \otimes U(1)_X$ is by now a nearly 30-year-old prediction \cite{FRA, PIPLEI}. In the absence of any reports of deviations from the Standard Model (SM), in particular by the LHC experimental collaborations, it is important to understand just how strong tensions between predictions from 331 Models and experimental results are. This is important on several grounds. 331 Models are based on the gauge symmetry $SU(3)_C \otimes SU(3)_L \otimes U(1)_X$ and as a $SU(N)$ type of group they predict the existence of $N^2 -1 = 8$ gauge bosons in its $SU(3)_L$ sector. Three of the gauge bosons are the familiar ones from the SM, namely the positively and negatively charged $W^{\pm}$ bosons and the neutral $Z^{0}$ boson. The other five non-SM gauge bosons are the positively and negatively double-charged bileptons $Y^{\pm\pm}$, the positively and negatively singly-charged bileptons $V^{\pm}$, and the heavy neutral $Z^{\prime}$ boson.

There are many interesting aspects of the 331 Models worth noticing \cite{MeiroseNepomuceno2011} but arguably, the most intriguing one is the explanation of three quark-lepton families, which is one of the main theoretical motivations for expecting bileptons in Nature. This is accomplished via a nontrivial anomaly cancellation in the model that takes place between families, unlike in the SM where the desired anomaly-free condition is accomplished for each family separately. 

Doubly-charged bileptons are the most striking prediction of 331 Models, but the expected LHC mass reach, considering the cleanest channel producing bileptons in proton-proton collisions, is $\sim$ 1 TeV and even if one considers the High-Luminosity LHC (HL-LHC) project, formerly known as Super-LHC (sLHC), the reach increases only by 20\% \cite{MeiroseNepomuceno2011}. Due to kinematics, however, these conclusions are heavily dependent on the leptoquarks masses that are predicted in the fermionic sector of the model. Should these particles be at least as heavy as 2 TeV the discovery reach for bileptons increases significantly \cite{Frampton2018}.
%It is worth mentioning that mass lower bounds on vector bileptons were established already 20 years ago. 
Regarding experimental constraints, a mass bound of  $M_Y >$ 740 GeV for doubly-charged bilepton gauge bosons was derived from constraints on fermion pair production at LEP and lepton-flavour violating charged lepton decays \cite{TUL}. It represented the most useful limit on doubly-charged vector bileptons for the past 20 years.  Searches for muonium-antimuonium conversion \cite{WILL} at PSI put the more stringent limit of $M_Y >$ 850 GeV on $Y^{\pm\pm}$, however, this is a less general limit as it assumes flavor-diagonal coupling for bileptons. Constraints on singly-charged bileptons are due to experimental limits on $\mu_{R} \to e \nu\nu$ and $\nu_\mu \to \nu_e$ oscillations, and are about 40\% lower than for doubly-charged bileptons \cite{Dion1998}. All these limits are now surpassed by the results of this article.
%For singly charged... \textcolor{red}{a bit tricky, will get back to it.}. Limits on $Z^{\prime}$.... ---> something strange is happening here @419

The most stringent limits on $Z^{\prime}$ gauge bosons within 331 Models up to now have been derived from weak charge data of Cesium and proton \cite{Long2018} and establish a lower bound of 4 TeV for these particles in the minimal version of the 331 Model, with a lowest value of 1.25 GeV for a version of the 331 Model discussed in \cite{Carcamo2017}. Recent limits from the ATLAS Collaboration imply a $Z^{\prime}$ mass lower limit of 5.1 TeV in the Sequential Standard Model \cite{Altarelli1989} at 95$\%$ confidence level. While this is the most stringent limit on any $Z^{\prime}$ mass, it is not applicable to 331 Models. As pointed out in Ref. \cite{Coriano2018}, the lower bounds of $Z^{\prime}$ mass can be significantly lower than those obtained from LHC, if other decay channels of $Z^{\prime}$ into new particles are included. The limits obtained in our work are, therefore, the most stringent limits for $Z^{\prime}$ in the 331 Model framework using hadron collision data.

This article is organized as follows. Section \ref{numerical} describes in detail the numerical implementation of the studied processes in Monte Carlo event generators and in a fast parametric detector simulation, including an overall description of the masses and couplings used in the simulations. In section \ref{cross} calculated cross-sections at 13 TeV center-of-mass energy, as well as widths for given particle masses are described. The limit setting procedures using a Bayesian approach is discussed in \ref{limits_valid}, followed by the specific limits obtained for $Z^{\prime}$, doubly-charged and singly-charged bileptons on sections \ref{limits_zprime}, \ref{limits_doubly} and \ref{limits_mono}, respectively. Conclusions are presented in section \ref{conclusion}.

% For the lower bounds on doubly charged vector bileptons still MY > 800 FeV  % https://arxiv.org/pdf/1812.02723.pdf --> 
% For the lower bounds on singly-charged vector bileptons still MY > 800 FeV  % https://arxiv.org/pdf/1812.02723.pdf -->
% For lower bounds on $Z^\prime$
% Also mention:
% bileptons have no color --> better suited to a linear collider

% We review this argument as follows.

%In the minimal version of the 331 Model, which uses minimal Higgs structure for spontaneous symmetry breaking, the $Z^\prime$ and the bilepton mass terms are coupled:

%\begin{equation}
%{M_Y \over M_{Z^\prime}} = {{\sqrt{3(1-4\sin^2\theta_W)} \over
%2\cos \theta_W }}
%\end{equation}

%This is an interesting relation, but aside from the resemblance it has with the $W$ and $Z^{0}$ mass relation in the SM, there is no compelling reason to believe it should be realized this way, so deviations from it are the rule rather than the exception.

% Regarding theoretical upper bounds in 331 Models 

%\section{The 331 Model}

\section{Numerical implementation}
\label{numerical}

\begin{figure}[ht]
\begin{subfigure}{.5\textwidth}
  \centering
  % include first image
  \includegraphics[width=.7\linewidth]{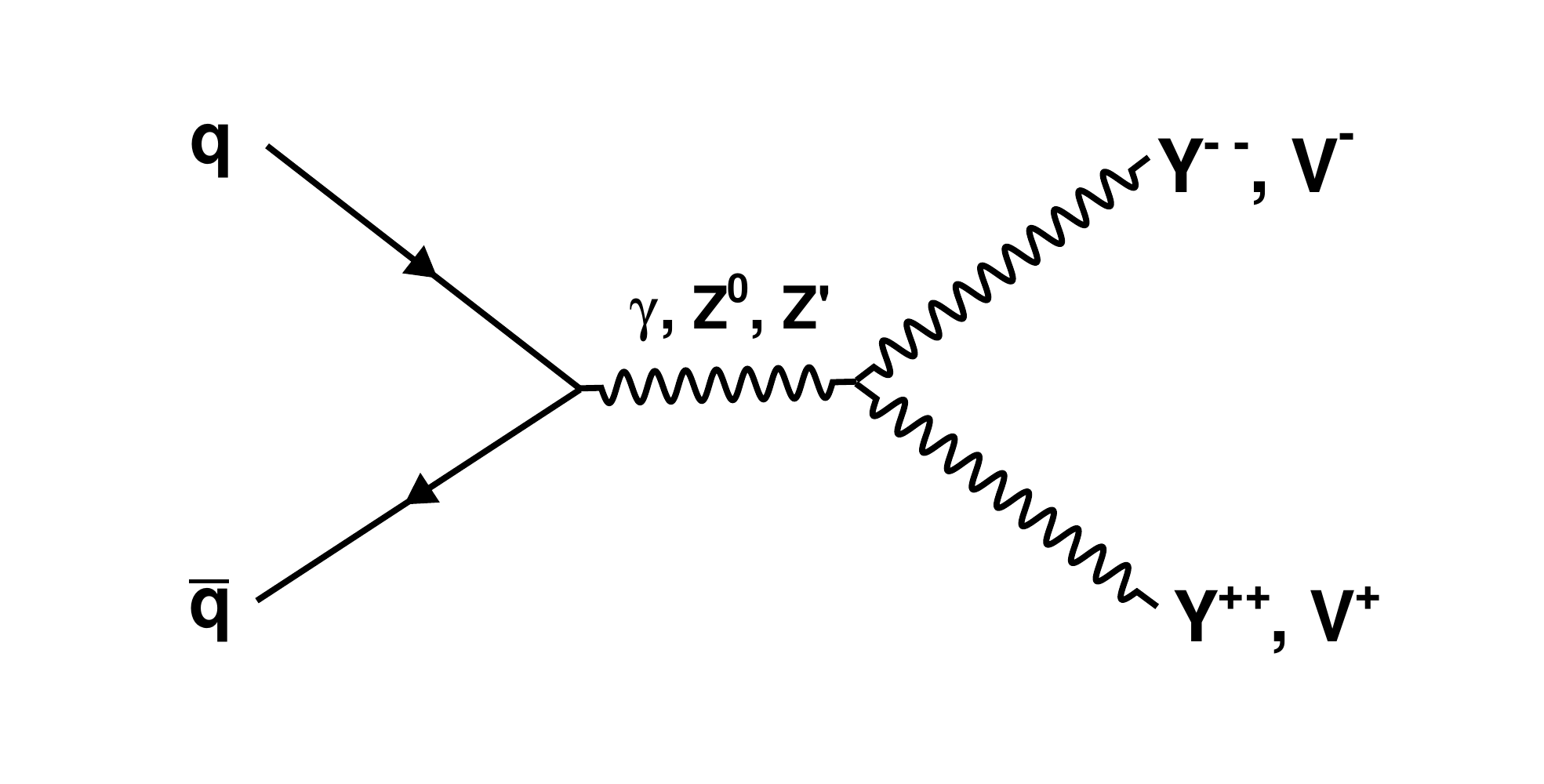}  
%  \caption{Put your sub-caption here}
  \label{fig:schannel}
\end{subfigure}
\begin{subfigure}{.5\textwidth}
  \centering
  % include second image
  \includegraphics[width=.7\linewidth]{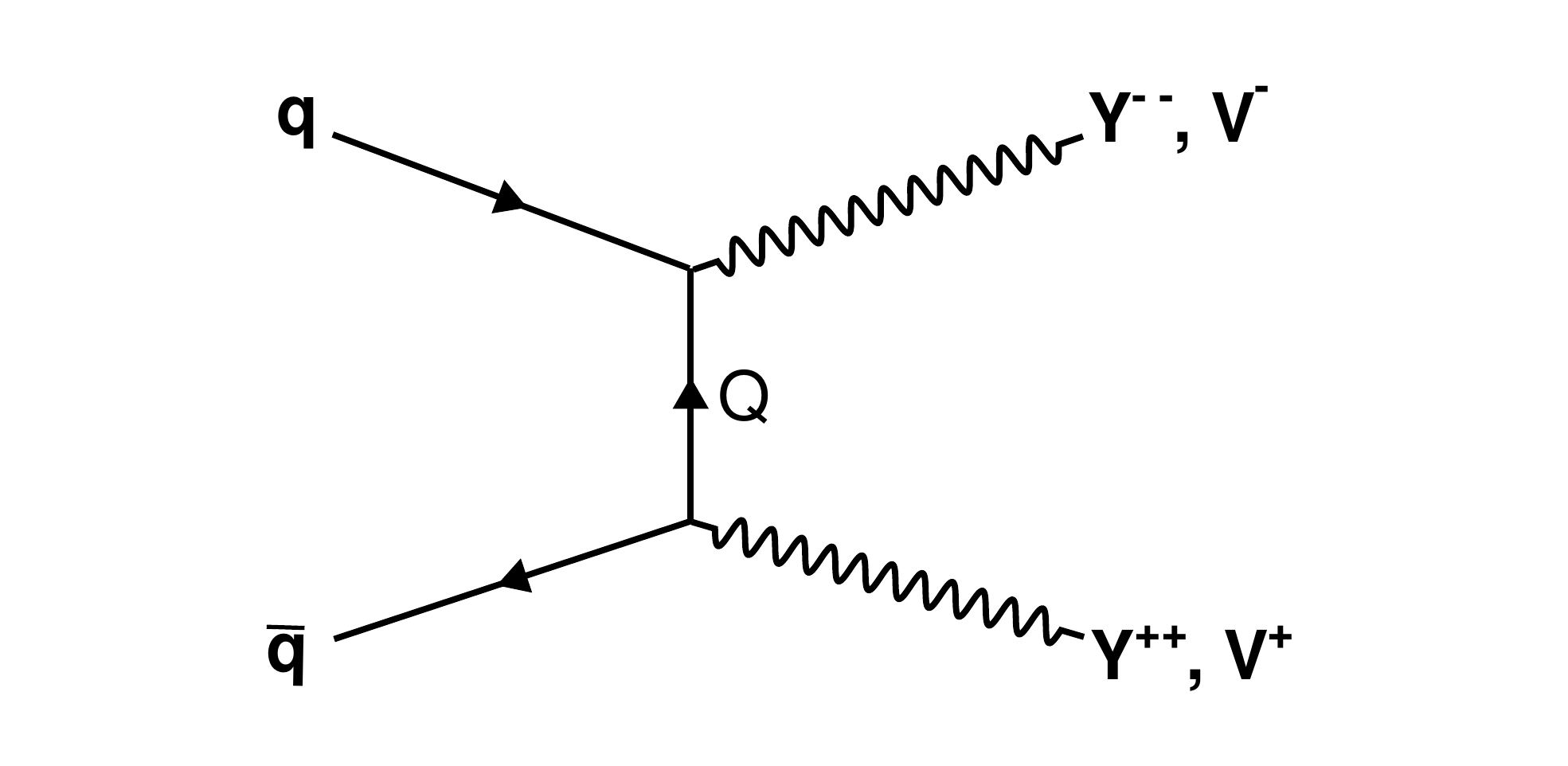}  
%  \caption{Put your sub-caption here}
  \label{fig:schannel}
\end{subfigure}
\caption{Feynman diagrams for bileptons production at LHC. The $t$-channel production is mediated by a leptoquark.}
\label{fig:feynman_diag}
\end{figure}

Bileptons can be produced in pairs at the LHC through a Drell-Yan-like process mediated by the photon, the SM $Z^{0}$ and the new neutral heavy boson $Z^{\prime}$. They can also be produced via a $t$-channel with a leptoquark exchange. The additional $t$-channels are needed in order to guarantee that all relevant quark sub-processes respect unitarity. The Feynman diagrams for bilepton production are shown in Figure \ref{fig:feynman_diag}.

\par
In doubly-charged bilepton pair production, each bilepton decay into a same-sign lepton pair, and therefore the natural processes to search for these type of bileptons is $pp \rightarrow Y^{++} Y^{--}\rightarrow \ell^{+}\ell^{+}\ell^{-}\ell^{-}X$, where $\ell \, = e,\mu$. For singly-charged bileptons, the characteristic decay is a signal with an opposite-charge lepton pair and two neutrinos, $pp \rightarrow V^{+} V^{-} \rightarrow \ell^{+}\ell^{-} \nu \nu X$. Leptonic decay is also the cleanest signal for $Z^{\prime}$ searches, hence in this work we investigate the processes $pp \rightarrow Z^{\prime} \rightarrow \ell^{+}\ell^{-} X$.

The production cross-sections for the 331 bosons were calculated using the \textsc{calchep} event generator \cite{belyaev2012}, where the model was previously implemented \cite{MeiroseNepomuceno2016}. Events were generated for all the processes mentioned above using the CTEQ6L1 parton distribution function and center-of-mass energy of 13 TeV. The generated events at parton level were processed by \textsc{pythia8} \cite{pythia8} to simulate parton shower, hadronization, and underlying event. A fast detector simulation was performed using \textsc{delphes} \cite{delphes} with ATLAS detector configuration. Pile-up is taken into account by overlaying minimum-bias events simulated with \textsc{pythia} with the hard-scatter process. The average number of $pp$ interaction per bunch crossing considered in this work is 24.

The bilepton masses were taken in the range of 400 GeV to 1300 GeV, and the $Z^{\prime}$ mass ranged from 1 TeV to 5 TeV. 
The leptoquarks masses were fixed at $M_{Q} =$ 1.5 TeV. When generating the bilepton events,  the $Z^{\prime}$ mass was fixed at 5 TeV. 
This particular choice of $M_{Z^{\prime}}$ has no significant impact on the bilepton cross-section, since the on-shell $Z^{\prime}$ exchange 
contribution is only relevant for $M_{Z^{\prime}} <$ 1 TeV. In order to perform a more general analysis, the coupling $g_{3l}$ between bileptons and leptons is also varied as a free parameter, following a similar approach used in Refs. \cite{Meirose2006, Meirose2008}. Besides the value $g_{3l} = 1.19\sqrt{4\pi\alpha}= 0.373$, that characterizes the 331 Model \cite{Frampton&Ng}, five other possibilities are investigated: $g_{3l} = $ 2, 4, 6, 8, 10. 

\section{Cross-sections and widths}
\label{cross}

\begin{figure}
\includegraphics[scale=0.4]{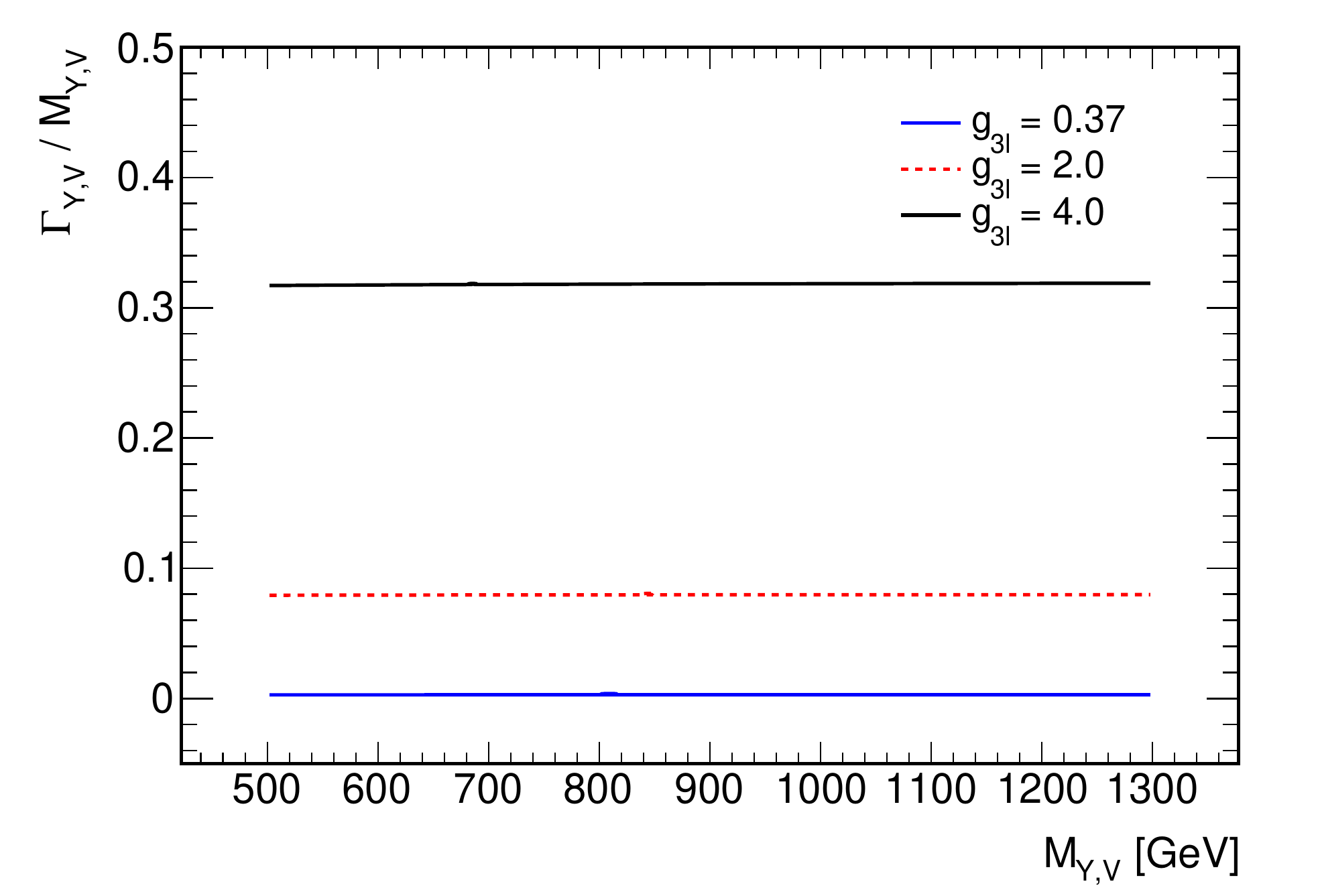}
\caption{\label{fig:bilepton_width} Width-mass ratio as a function of mass for bileptons calculated for three values of $g_{3l}$.}
\end{figure}

The width-mass ratio ($\Gamma/M$) for bileptons is shown on Figure \ref{fig:bilepton_width}. The ratio for three representative values of $g_{3l}$ are plotted. For small values of $g_{3l}$, the bilepton resonance is very narrow, with $\Gamma/M < $ 1\%. For $g_{3l}>$ 2, it increases rapidly, reaching $\sim$ 30\% for $g_{3l}$ = 4 and $\sim$ 190\% for $g_{3l}$ = 10. It can be noticed that, since the bilepton decays are the same for the interval shown, the ratio remains constant for each value of $g_{3l}$.

\par
\begin{figure}
\includegraphics[scale=0.4]{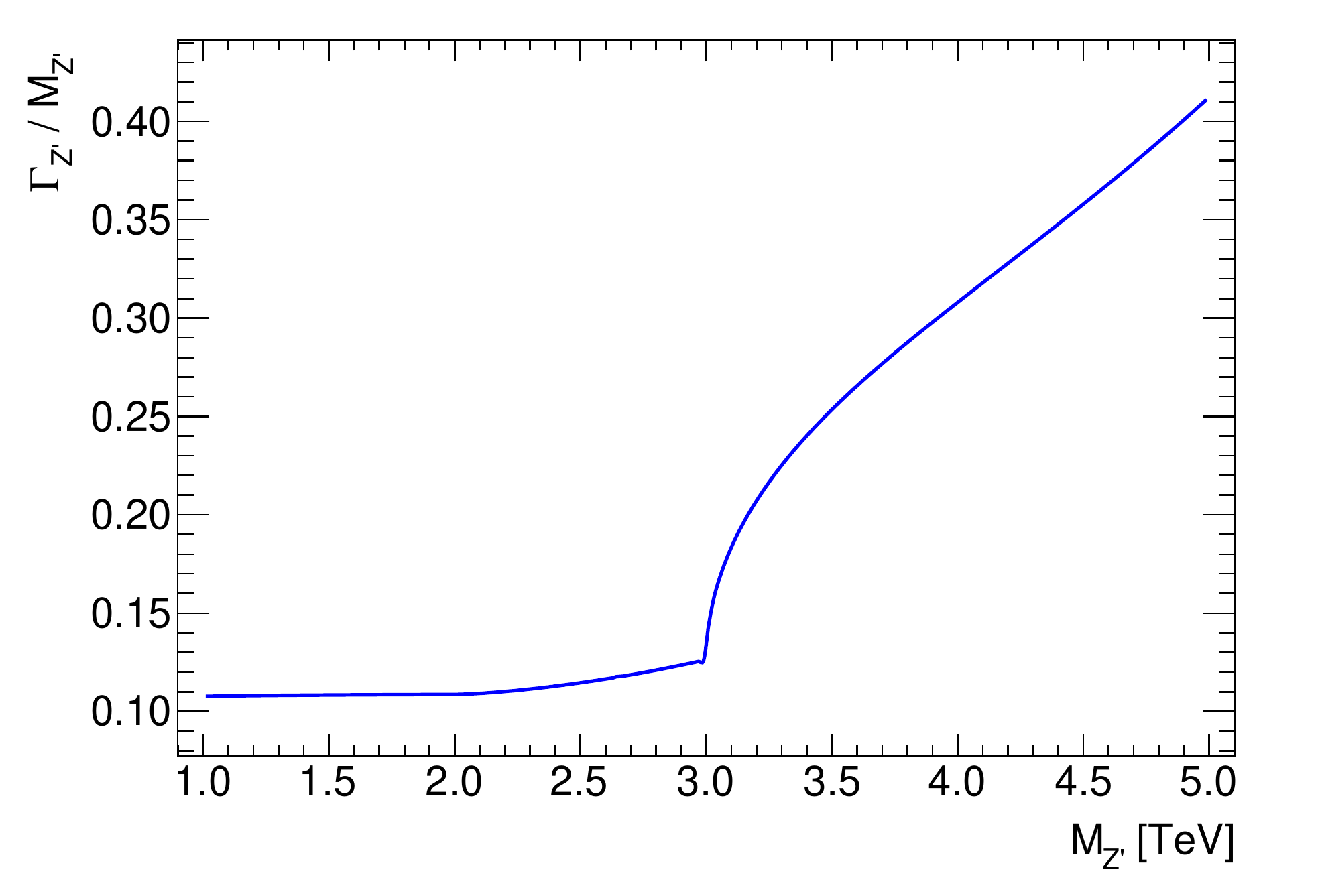}
\caption{\label{fig:zprime_width} Width-mass ratio as a function of $Z^\prime_{331}$ mass. From $M_{Z^\prime_{331}} =$ 3 TeV, the decay $Z^\prime_{331} \rightarrow Q\bar Q$ becomes kinematically allowed. }
\end{figure}

Figure \ref{fig:zprime_width} shows the $\Gamma/M$ ratio for  $Z^{\prime}_{331}$. The $Z^{\prime}$ is a broad resonance in the minimal 331 Model. The $Z^{\prime}_{331}$ width becomes larger from $M_{Z^{\prime}_{331}}$ = 3 TeV as a result of the boson decay to the model's leptoquarks.  

\par

\begin{figure}
\includegraphics[scale=0.4]{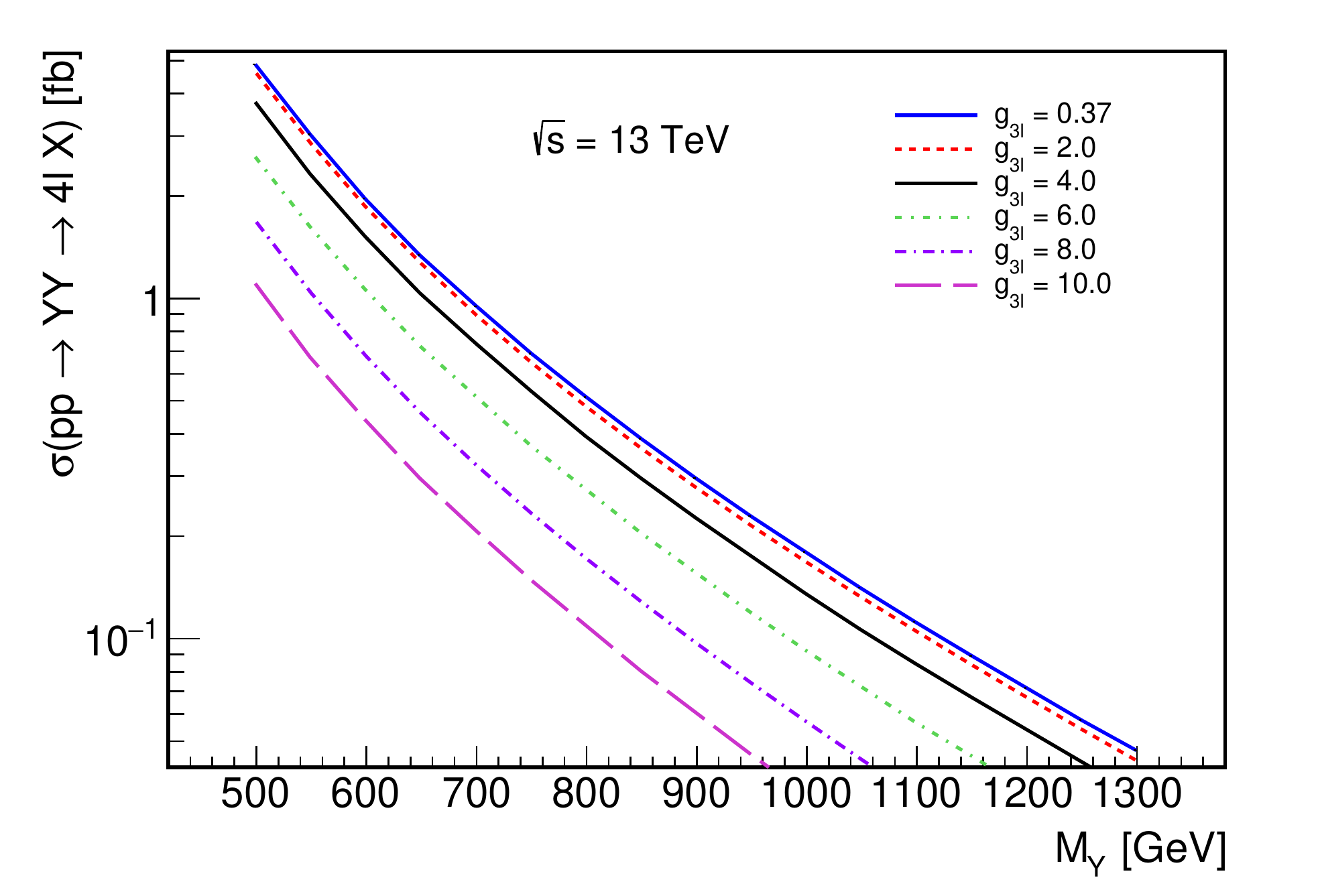}
\caption{\label{fig:xsec} LO cross-section for four leptons production mediated by doubly or singly bileptons considering different values of $g_{3l}$ at $\sqrt{s} =$ 13 TeV.}
\end{figure}

Figure \ref{fig:xsec} shows the cross-section for four-lepton production mediated by bileptons as a function of $M_{Y}$, for different coupling values.
The production of bilepton pairs does not depend on $g_{3l}$, and the branching fractions $Br(Y^{\pm \pm} \rightarrow \ell^{\pm} \ell^{\pm})$ are not affected by $g_{3l}$ variation. 
The total bilepton width, however, increases as $g_{3l}$ rises, causing the total cross-section to decrease.

This cross-section behavior is characteristic for the case we are studying, where the bileptons decay only to leptons and not to 
leptoquarks ($M_{Q} > M_{Y,V}$). The cross-section for the process $pp \rightarrow V^{+} V^{-} \rightarrow \ell^{+}\ell^{-} \nu \nu X$ is the same as for $pp \rightarrow Y^{++} Y^{--}\rightarrow \ell^{+}\ell^{+}\ell^{-}\ell^{-}X$. The LO cross-sections are multiplied by a $k$-factor of 1.25 to take into account NLO effects \cite{muhlleitner2003}.
\par
The di-lepton production cross-section mediated by $Z^\prime_{331}$ is shown by the solid red line of Figure \ref{fig:zprime_limit}. The small bump around 3 TeV indicates the point from where the exotic decay $Z^\prime_{331} \rightarrow Q\bar Q$ becomes kinematically allowed. 

%\subsection{Lepton-pair production plus $\met$ mediated by $V^{\pm}$}

%\subsection{Lepton-pair production mediated by $Z^\prime$}

\section{Limit Setting Procedure and Validation}
\label{limits_valid}

The invariant mass distributions $m ({\ell \ell})$ of lepton pairs derived from doubly-charged bileptons and from ${Z^{\prime}}$ productions are used to calculate upper limits on $\sigma B$, where $\sigma$ is the cross-section of the new physics process and $B$ is the branching ratio of 
${Z^{\prime}}$ and $Y^{\pm \pm}$ decaying to leptons. For the singly-charged bilepton analysis, another distribution, the so-called "stransverse" mass \cite{lester1999, lester2014}, is used. A likelihood function defined as the product of Poisson probabilities over all distribution's bins of a given search is 
constructed. To calculate the limits, a Bayesian approach is applied with a flat prior probability distribution for $\sigma B$, 
and the Bayes theorem is employed to evaluate the marginal posterior probability density function, $\mathcal{L}(\sigma B | N)$, 
where $N$ is the number of observed events. Upper limits at 95\% CL are set on $\sigma B$ by integrating the posterior probability density function as

\begin{equation}
0.95 = \frac{\int_{0}^{(\sigma B)_{up}} \mathcal{L}(\sigma B | N) d(\sigma B)}{\int_{0}^{\infty}\mathcal{L}(\sigma B | N) d(\sigma B)}
\end{equation} 

\noindent
\newline
where $(\sigma B)_{up}$ is the calculated limit. The calculation is performed with the Bayesian Analysis Tool Kit \cite{bat2009}.
The upper limits on $\sigma B$ are translated into lower limits on the mass of the new bosons by using the theoretical cross-section 
for the new boson production. The limits obtained with data are called observed limits. The expected limits are obtained by 
running a large number of pseudo-experiments where it is assumed that only background events are present. 
All the estimated backgrounds and efficiencies used in our analysis are extracted from the relevant ATLAS publications. 

\par
In order to validate the signals' simulation and the limit setting procedure described above, we calculate limits on $Z^{\prime}_{SSM}$ and $Z^{\prime}_{\chi}$ models, with the $Z^{\prime}$ signals simulated with 
\textsc{pythia} and \textsc{delphes}, and compare the results with those obtained by the ATLAS Collaboration \cite{atlaszprime2017}.  
Table \ref{Tab:table1} show the results for electron and muon channels combined. As we can see, there is a good agreement between the ATLAS 
results and the limits obtained using the signals simulated with \textsc{delphes}.

\begin{table}
\renewcommand{\tabcolsep}{2.2mm}
\caption{Comparison between the limits on two popular $Z^\prime$ models obtained by ATLAS Collaboration and by the \textsc{pythia} and \textsc{delphes} simulation performed in this work.}
\label{Tab:table1}
\begin{center}
\begin{tabular}{|c|cc|cc|}
\hline
\multirow{2}{1.5cm}{Model}& \multicolumn{4}{c|}{\centering Lower Limits on $M_{Z^{\prime}}$ [TeV] } \bigstrut \\
\cline{2-5} \hspace{2cm} & \multicolumn{2}{c|}{\centering ATLAS }  & \multicolumn{2}{c|}{DELPHES} \bigstrut  \\ 
\cline{2-5}
{} & obs. & exp. & obs. & exp. \bigstrut \\
\hline
$Z^{\prime}_{SSM}$  & 4.5 & 4.5 & 4.4 & 4.4 \bigstrut \\
$Z^{\prime}_{\chi}$ & 4.1 & 4.0 & 4.0 & 4.0 \bigstrut \\
\hline
\end{tabular}
\end{center}
\end{table}

To set limits on bileptons masses an couplings, and on $Z^{\prime}$ mass, different ATLAS searches are considered.
The data sample in all the analysis corresponds to an integrated luminosity of 36.1 fb$^{-1}$. These are discussed in the following (\ref{limits_zprime}-\ref{limits_doubly}) sections.

\section{Limits on $Z^\prime_{331}$} 
\label{limits_zprime}

Limits on $Z^\prime_{331}$ mass are set using the data from Ref. \cite{atlaszprime2017}. 
The signal candidates are selected by requiring at least one pair of same flavour leptons (electrons or muons). 
Electrons candidates are selected if they have transverse energy ($E_T$) greater than 30 GeV and pseudorapidity $|\eta| <$ 2.47. 
Events inside the region 1.37 $\leq |\eta| \leq$ 1.52 are excluded due to poor energy resolution. 
Muons candidates are required to have transverse momentum ($p_T$) greater than 30 GeV, $|\eta| <$ 2.5 and opposite charges. 
The opposite charge requirement is not applied to the electron channel since it is not applied in data. 
If more than two leptons are found, the ones with highest transverse momentum are kept.
\par

\begin{figure}[ht]
\begin{subfigure}{.5\textwidth}
  \centering
  % include first image
  \includegraphics[width=.9\linewidth]{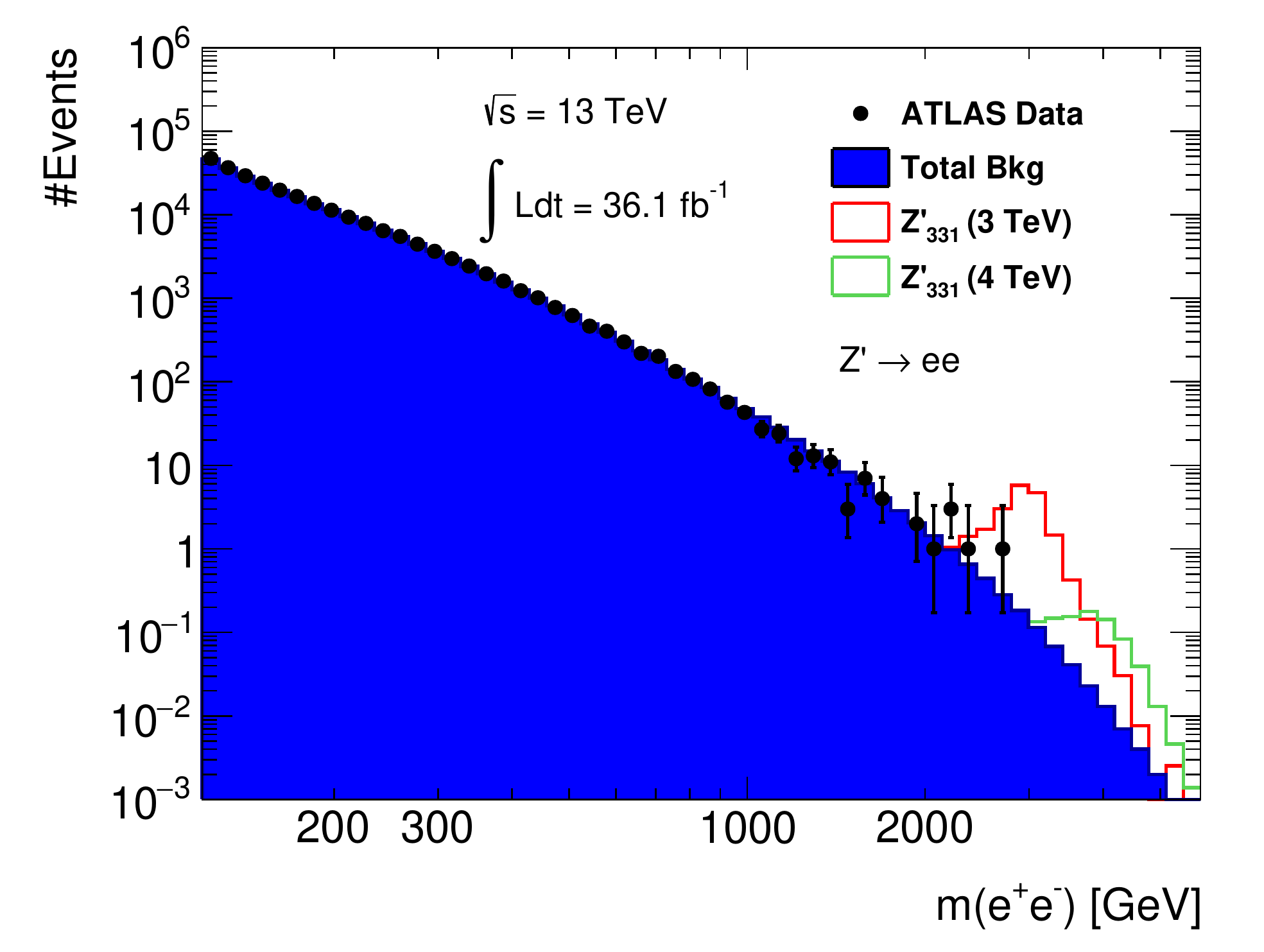}  
  \caption{}
  \label{fig:mass_ee}
\end{subfigure}
\begin{subfigure}{.5\textwidth}
  \centering
  % include second image
  \includegraphics[width=.9\linewidth]{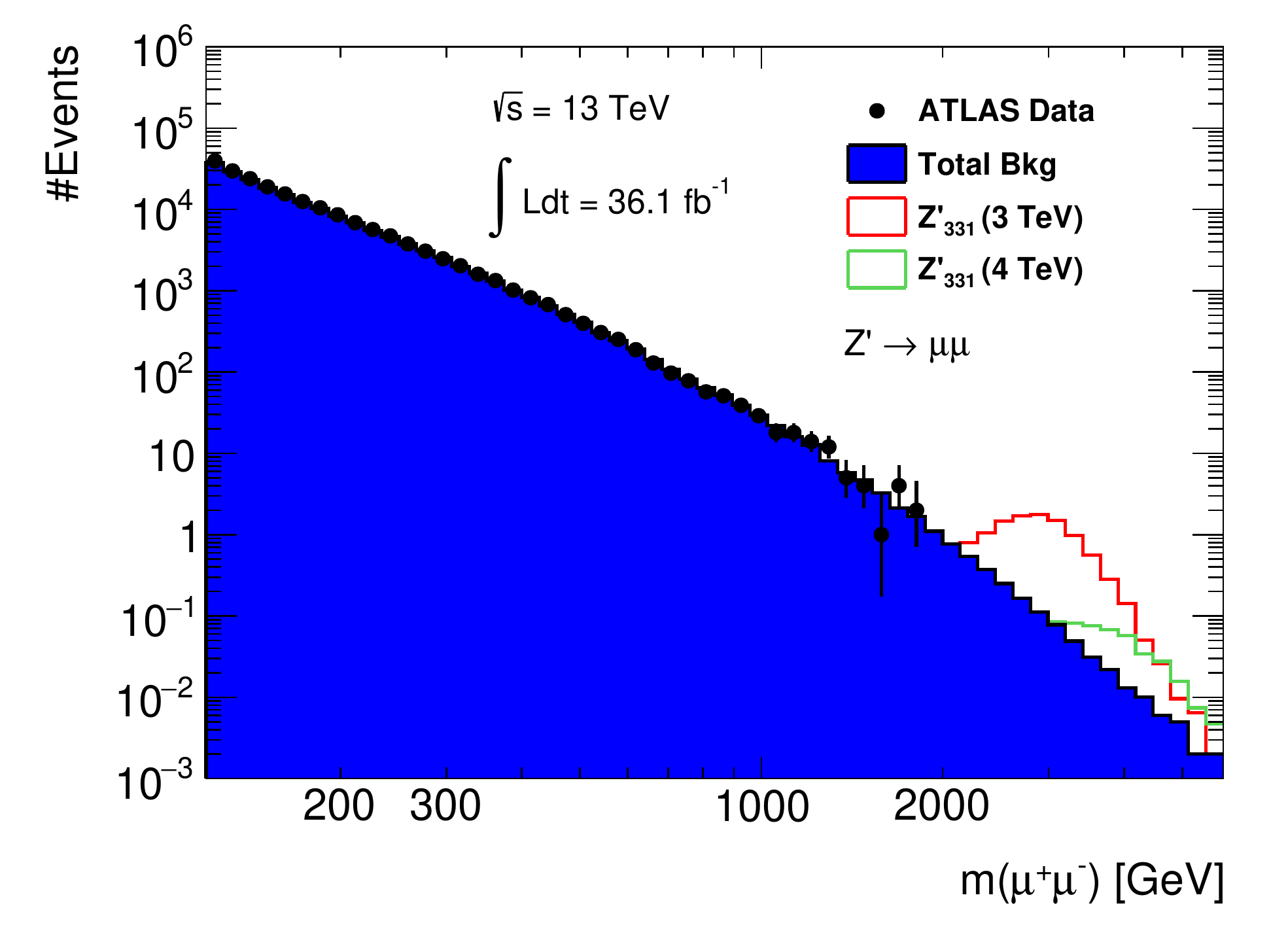}  
  \caption{}
  \label{fig:mas_mm}
\end{subfigure}
\caption{di-lepton invariant mass distribution for electron (a) and muon (b) channels. The black points show the ATLAS data with the statistical uncertainties. The dark histogram is the total background, and the red and green open histograms are the distributions for two $Z^{\prime}_{331}$ mass hypotheses. }
\label{fig:zprime_mass}
\end{figure}

The inputs to the statistical analysis are the reconstructed invariant mass distribution of the selected events shown in Figure \ref{fig:mass_ee} and Figure \ref{fig:mas_mm} for electron and muon channels, respectively. The data points are shown with their statistical uncertainties. The dark histogram is the total estimated background. Two $Z^{\prime}_{331}$ signals with masses of 3 TeV and 4 TeV are also displayed for comparison.

\begin{figure}
\includegraphics[scale=0.4]{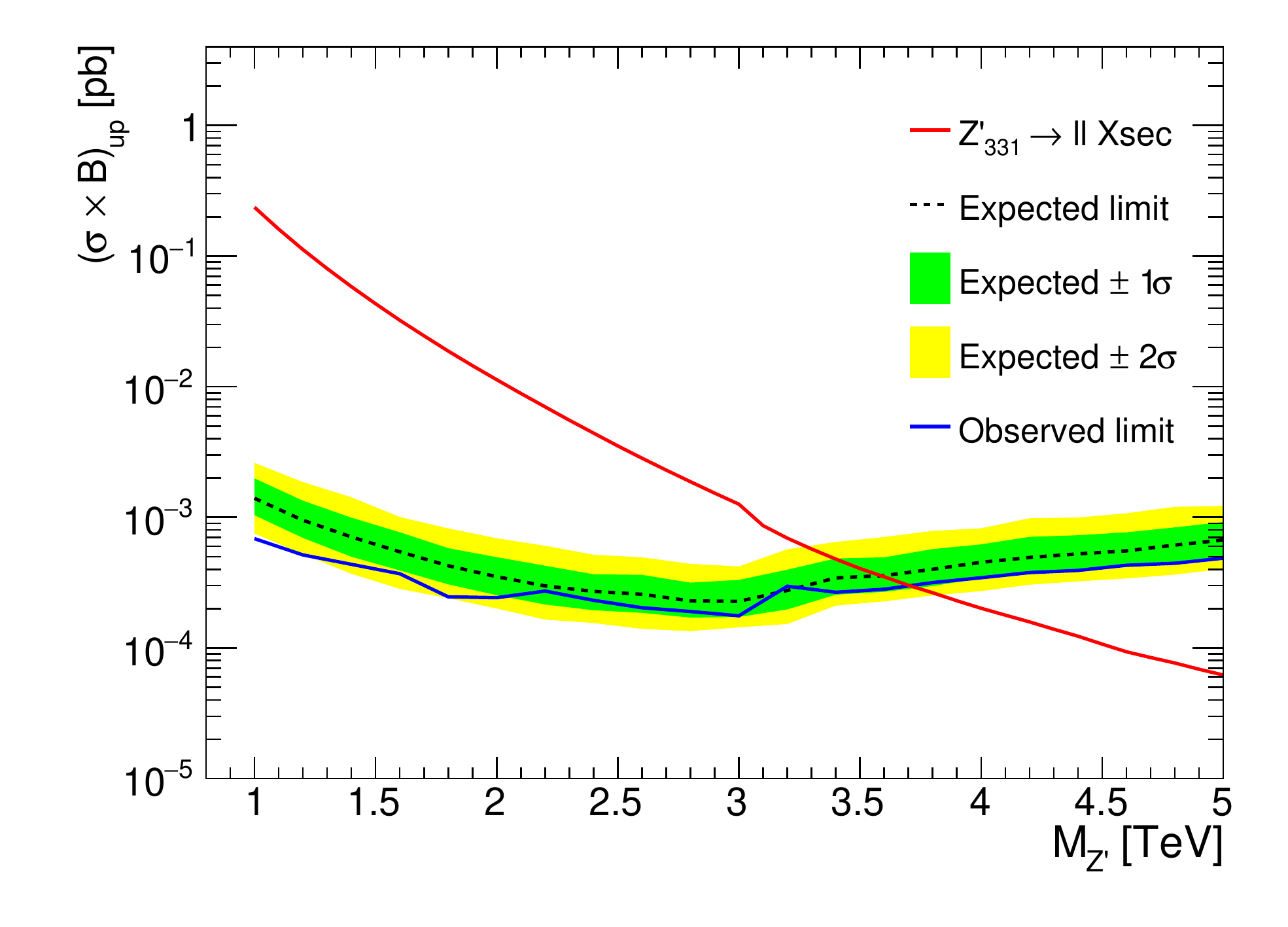}
\caption{\label{fig:zprime_limit} Observed and expected limits on $\sigma B$ as a function of the $Z^{\prime}_{331}$ mass assumption. The green and yellow bands show the $1\sigma$ and $2\sigma$ variation on the expected limit, respectively. The solid red line is the theoretical cross-section for di-lepton production mediated by $Z^{\prime}_{331}$. }
\end{figure}

The observed and expected upper limits $(\sigma B)_{up}$ are calculated for various $Z^{\prime}_{331}$ mass hypothesis and the results are shown in Figure \ref{fig:zprime_limit}. The lower limits on $M_{Z^{\prime}_{331}}$ are extracted from the crossing point between the theoretical 
cross-section and $(\sigma B)_{up}$.  
The observed and expected mass limits found are 3.7 TeV and 3.6$^{+0.1}_{-0.2}$ TeV, respectively. 
These limits represent significant improvement compared to previous bounds on $Z^{\prime}$ from 331 Models \cite{salazar2015,farinaldo2013,coutinho2013}.

\section{Limits on Doubly Charged Bileptons}
\label{limits_doubly}

In this section, the ATLAS search for doubly charged Higgs bosons \cite{atlasdoubly2018} are interpreted in terms of doubly charged bileptons. The analysis focuses on $Y^{\pm \pm}\rightarrow e^{\pm}e^{\pm}$ and $Y^{\pm \pm}\rightarrow \mu^{\pm}\mu^{\pm}$ decay channels. The bilepton decay in different lepton flavors,
$Y^{\pm \pm}\rightarrow e^{\pm}\mu^{\pm}$, is not considered since there is no usable data available for this particular channel. 
\par
As the doubly charged bileptons are produced in pairs, the selected final state must have at least three same flavour leptons, all within the inner detector coverage ($|\eta| < 2.5$). Events with three leptons are required to have one same-sign lepton pair and one lepton of opposite charge ($\ell^{\pm}$ $\ell^{\pm}$ $\ell^{\mp}$). If there are four leptons in the event, the net electric charge must be zero. Following the same selections applied to data, events with at least one $b$-tagged jet are rejected to suppress background from top quarks. An event is also rejected if the invariant mass of the opposite charge same flavor leptons pair is in the range 82.1 GeV $< m(\ell^{+}\ell^{-}) <$ 101.2 GeV. In the three leptons events, signal sensitivity is optimized by imposing the same charge lepton separation to be $\Delta R(\ell^{\pm}\ell^{\pm}) >$  3.5 and their combined transverse momentum to be $p_{T}(\ell^{\pm}\ell^{\pm}) >$ 100 GeV. In addition, the scalar sum of the individual leptons transverse momenta is required to be greater than 300 GeV. 
\par
The invariant mass of the same sign leptons pair is used as discriminant variable. In the case of events with four leptons, the variable considered is the average invariant mass of the two same charge leptons: 

\begin{equation}
\overline{m} \, \equiv \, \frac{m(\ell^{+}\ell^{+}) + m(\ell^{-}\ell^{-})}{2}.
\end{equation} 

\begin{figure}[ht]
\begin{subfigure}{.5\textwidth}
  \centering
  % include first image
  \includegraphics[width=.9\linewidth]{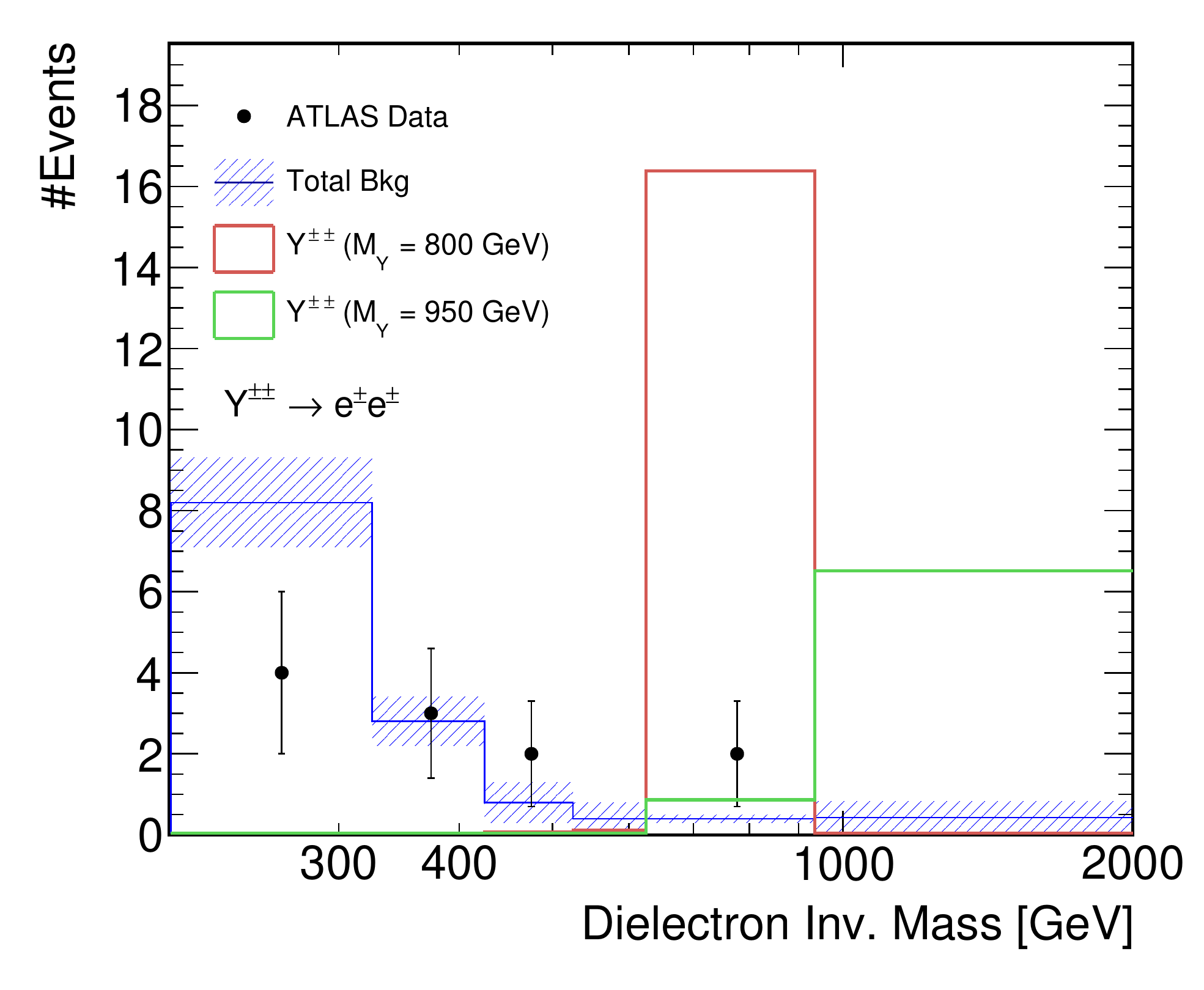}  
  \caption{}
  \label{fig:mass_dielec}
\end{subfigure}
\begin{subfigure}{.5\textwidth}
  \centering
  % include second image
  \includegraphics[width=.9\linewidth]{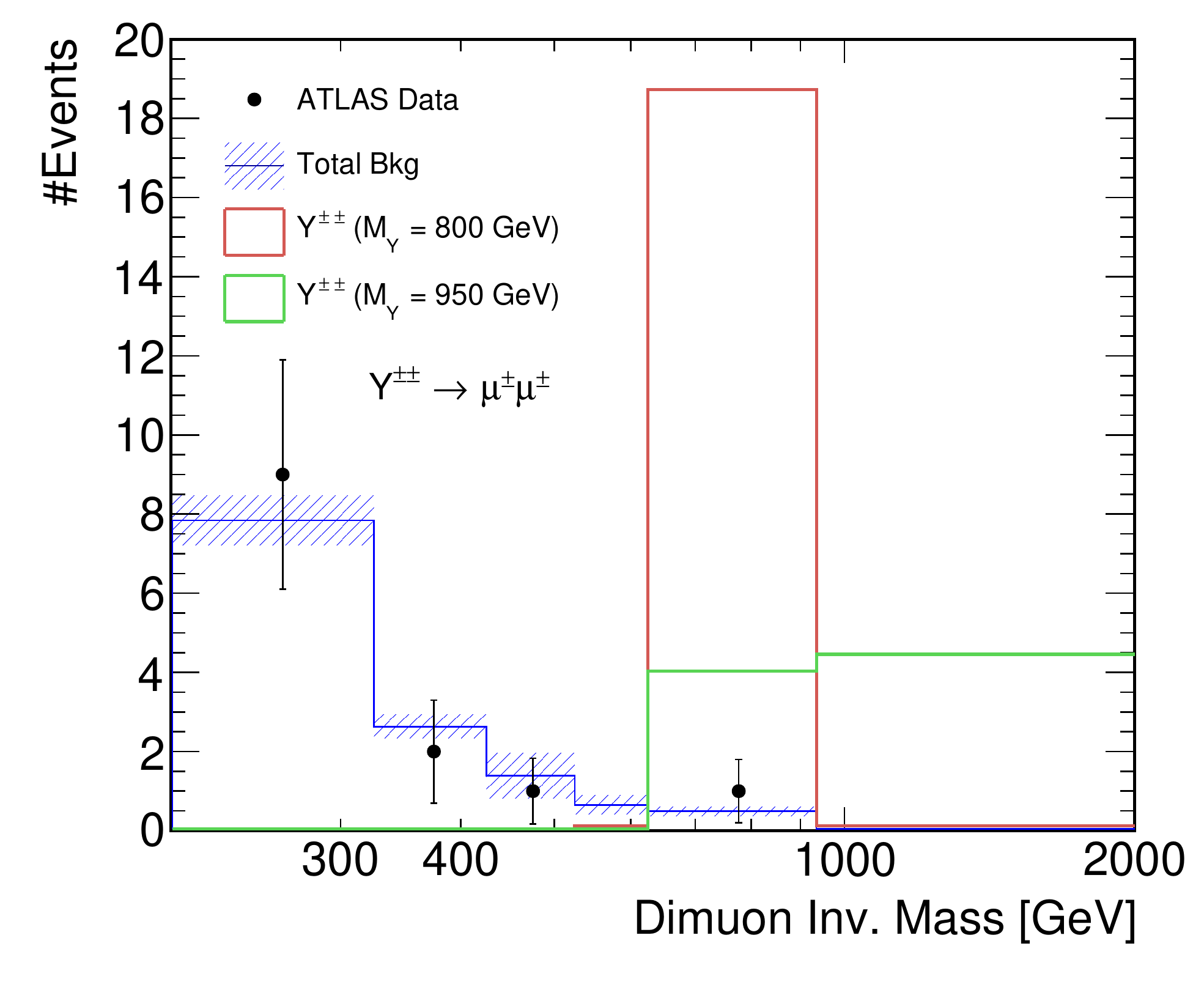}  
  \caption{}
  \label{fig:mass_dimuon}
\end{subfigure}
\caption{Invariant mass distribution for same-charge electrons (a) and same-charge muons (b) pairs. The hatched band represents the total systematic uncertainty on the background determined by ATLAS. Two doubly-bileptons signals of masses 800 GeV and 950 GeV are shown.  }
\label{fig:bilepton_mass}
\end{figure}

The analysis is performed in the region where $m(\ell^{\pm}\ell^{\pm})$ and $\overline{m}$ are both above 200 GeV. Figures \ref{fig:mass_dielec} and \ref{fig:mass_dimuon} show the invariant mass for three and four leptons events combined, in electron and muon channels. Along with the ATLAS data and background, two bilepton signals are shown. The hatched bands represent the systematic uncertainty on the background. 

\par
\begin{figure}
\includegraphics[scale=0.45]{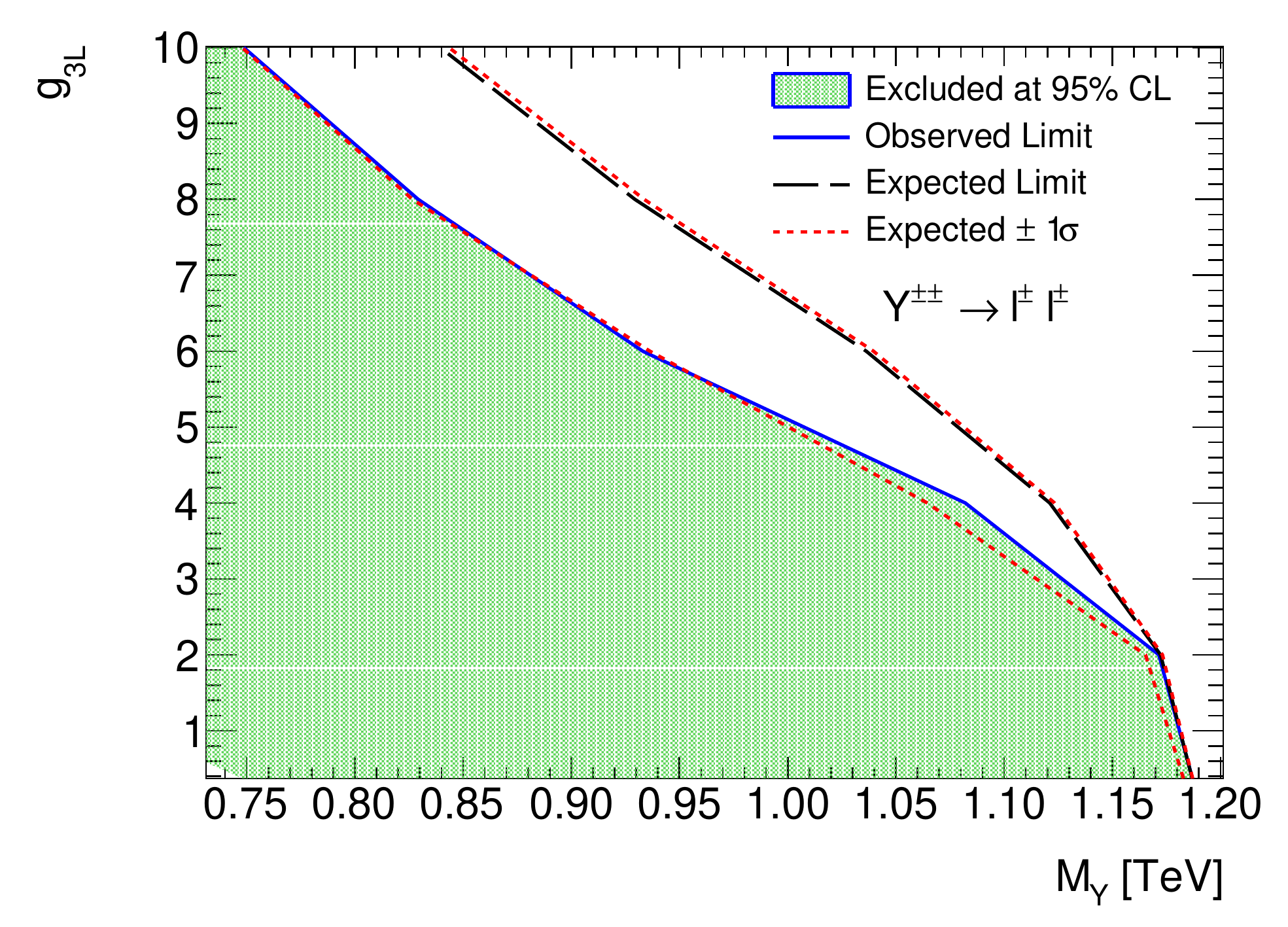}
\caption{\label{fig:doubly_limit} Excluded parameter space for doubly-charged bileptons. The observed and expected limits are shown by the solid blue and long-dashed lines, respectively. The dashed red lines represent the $1\sigma$ variation on the expected limits.}
\end{figure}

Limits are set considering different bilepton mass and coupling ($g_{3l}$) hypothesis. The excluded parameter space is shown in Figure \ref{fig:doubly_limit}, from where we conclude that doubly-charged bileptons with masses between $\sim$ 750 GeV and $\sim$ 1200 GeV are excluded. For high values of $g_{3l}$, the observed limits are weaker than the expected due to the larger bilepton width. It it interesting to note that the four leptons production cross-section does not change significantly for $g_{3l} <\,$ 1, which means that $M_{Y} > \,$ 1.2 TeV is the maximum bound on doubly-charged bileptons mass that can be reached with the data considered. 

\section{Limits on singly-charged Bileptons}
\label{limits_mono}

The production of singly-charged bileptons are characterized by events with two leptons and missing transverse energy in the final state. It is the same measurable final state studied in charginos and sleptons searches from supersimetric models. Here we use the ATLAS data from supersimetric searches \cite{atlassusy2018} to set bounds on singly-charged bileptons. This analysis considers all possible leptonic final states produced by the singly-charge bilepton pair production: $V^{\pm}\rightarrow e^{+} \nu e^{-}\nu$, $V^{\pm}\rightarrow \mu^{+}\nu \mu^{-}\nu$, $V^{\pm}\rightarrow e^{+}\nu \mu^{-}\nu$ and $V^{\pm}\rightarrow e^{-}\nu \mu^{+}\nu$.
\par
The signal events candidates are required to have exactly two leptons and missing transverse energy. The leptons can be of the same or different flavors, but they must have opposite charges. Additionally, leptons are required to have $|\eta| <$ 2.5 and $p_{T} >$ 10 GeV, and a di-lepton invariant mass of $m_{\ell \ell} >$ 110 GeV. An event is rejected if there is a $b$-tagged jet with $p_{T} >$ 20 GeV or any other jet with $p_{T} >$ 60 GeV. 
The selected events are separated in two categories: same-flavor ($e^{+}e^{-}$ and $\mu^{+}\mu^{-}$ events) and different-flavor ($e^{\pm}\mu^{\mp}$ events).
\par
The final selection applied is based on the the variable ``stransverse mass'', $m_{T2}$, defined as \cite{lester1999, lester2014}:

\begin{equation}
m_{T2} \; \equiv \; \underset{q_{T}}{\min} \Bigl[ \max \Bigl( m_T(\textbf{p}^{\ell 1}_{T}, \textbf{q}_T), m_{T}(\textbf{p}^{\ell 2}_{T}, \textbf{p}^{\text{miss}} - \, \textbf{q}_T) \Bigr )\Bigr],
\end{equation} 

\noindent
where 
\begin{equation}
m_{T}(\textbf{p}_{T}, \textbf{q}_T) \, = \sqrt{2 (p_{T} q_{T} - \textbf{p}_T . \textbf{q}_T)},
\end{equation}  

\noindent
$\textbf{p}^{\ell 1}_{T}$ and $\textbf{p}^{\ell 2}_{T}$ are the two leptons vectors transverse momentum,  $\textbf{p}^{\text{miss}}$ is the missing transverse momentum and $\textbf{q}_T$ is a transverse momentum vector that minimizes the larger of $m_T(\textbf{p}^{\ell 1}_{T}, \textbf{q}_T)$ and $m_{T}(\textbf{p}^{\ell 2}_{T}, \textbf{p}^{\text{miss}} - \, \textbf{q}_T)$.
\par
\begin{figure}[ht]
\begin{subfigure}{.5\textwidth}
  \centering
  % include first image
  \includegraphics[width=.9\linewidth]{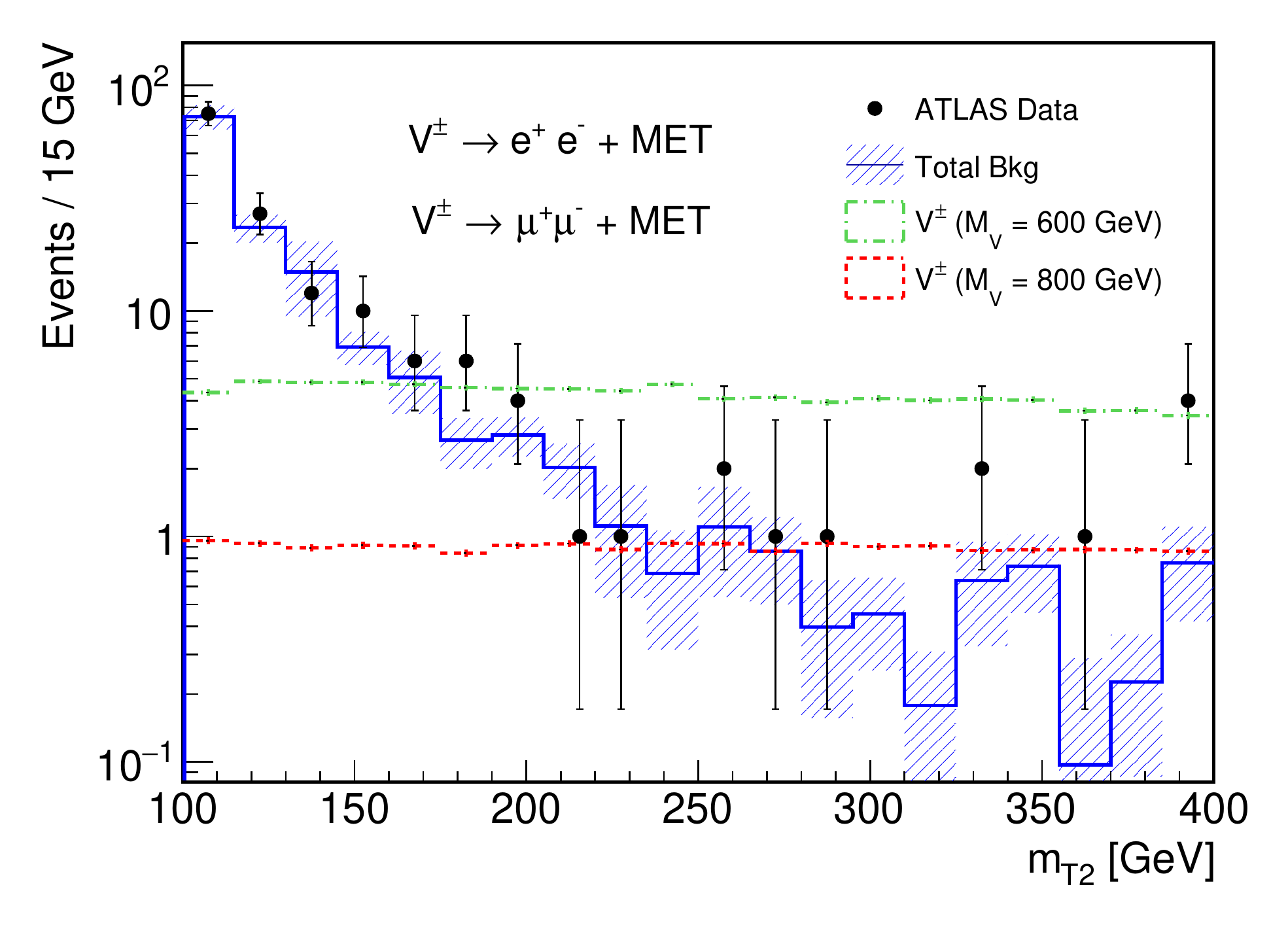}  
  \caption{}
  \label{fig:mt2_sf}
\end{subfigure}
\begin{subfigure}{.5\textwidth}
  \centering
  % include second image
  \includegraphics[width=.9\linewidth]{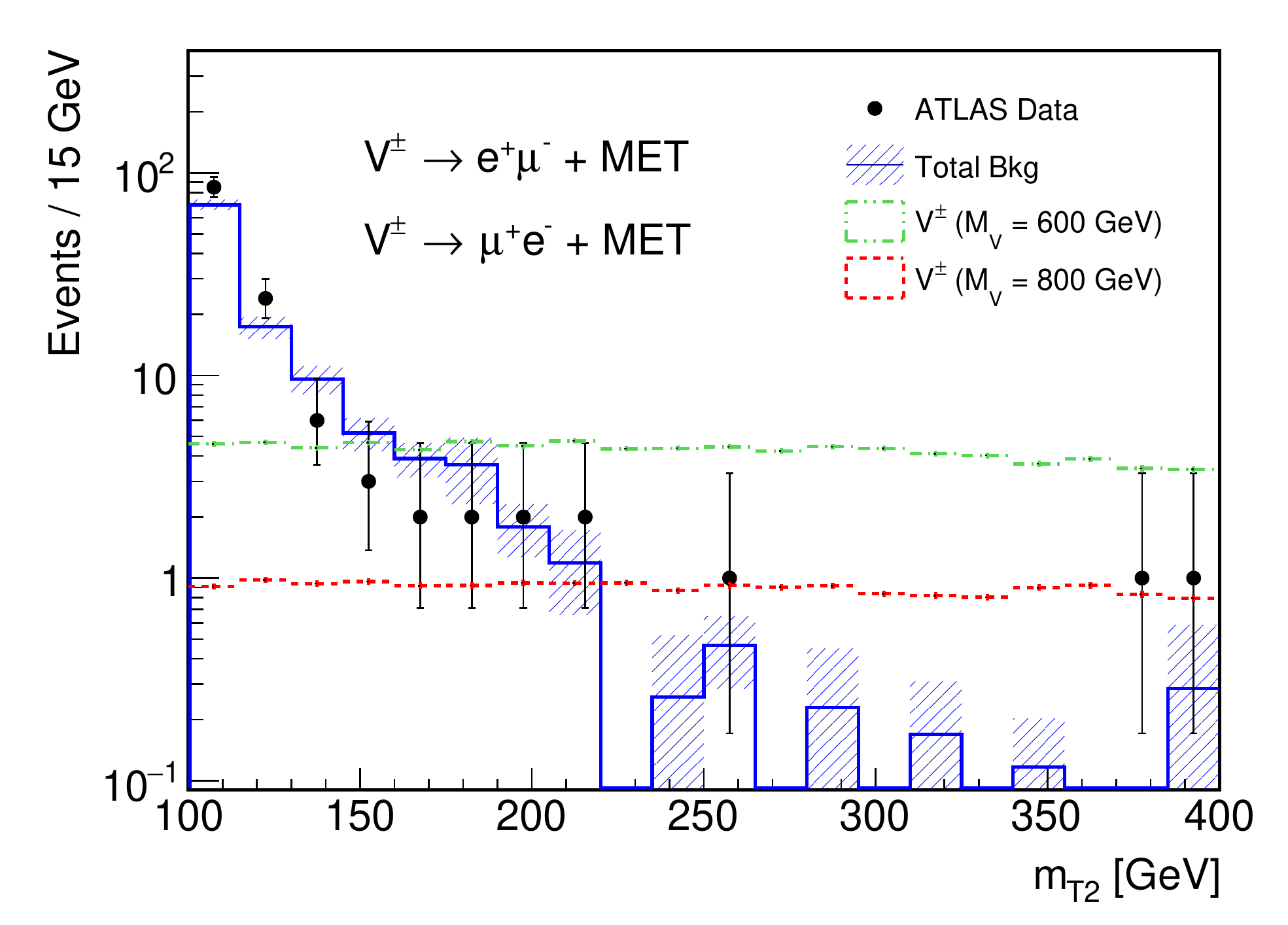}  
  \caption{}
  \label{fig:mt2_df}
\end{subfigure}
\caption{$m_{T2}$ distribution for ATLAS data, total background and two simulated singly-charged bilepton signals. The same flavor $m_{T2}$ distributions (a) includes $e^{+}e^{-}$ and $\mu^{+}\mu^{-}$ events, while the different flavor (b) corresponds to $e^{\pm}\mu^{\mp}$ events. }
\label{fig:singly_mass}
\end{figure}

The $m_{T2}$ variable is used in the statistical analysis where the signal region is defined by requiring $m_{T2} >$ 100 GeV. Figures \ref{fig:mt2_sf} and \ref{fig:mt2_df} show the $m_{T2}$ distribution for the selected events in the same flavor and different flavor categories, respectively. 
Two signal examples are also shown. 

\par
\begin{figure}
\includegraphics[scale=0.45]{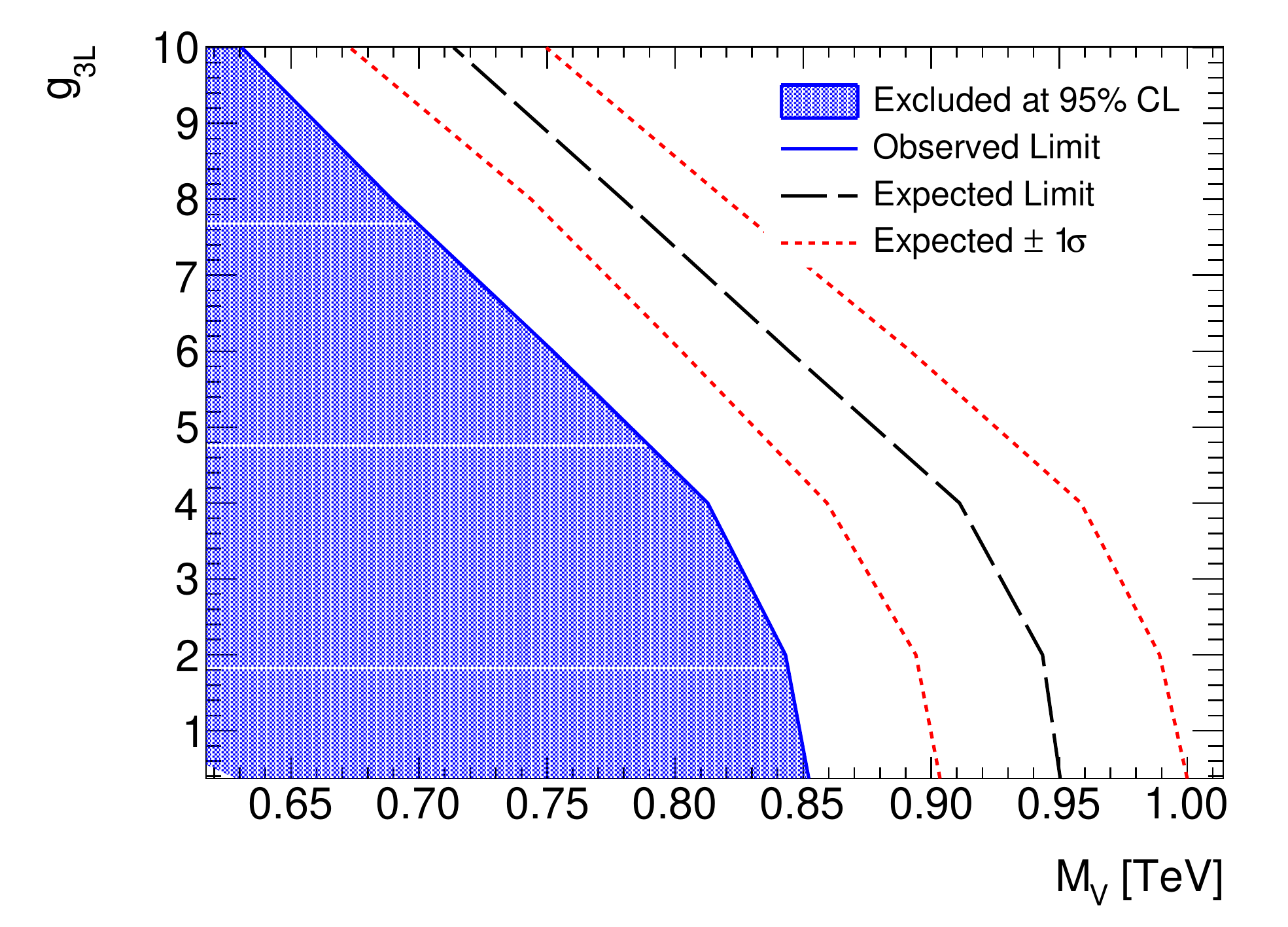}
\caption{\label{fig:mono_limit} Excluded parameter space for singly-charged bileptons. The observed and expected limits are indicated. The dashed red lines represent the $1\sigma$ variation on the expected limits.}
\end{figure}

The observed and expected exclusion limits are displayed in Figure \ref{fig:mono_limit}. Singly-charged bileptons with masses between 600 GeV and 850 GeV are excluded at 95\% CL. The observed limits are smaller than the expected ones due to the data excess observed in the region $m_{T2} >$ 250 GeV, as we can see in Figure \ref{fig:singly_mass}. These results are the first limits derived for singly-charged bileptons using LHC data.

\section{Conclusions}
\label{conclusion}

Limits on the mass and couplings of vector bosons from 331 Models are derived using different ATLAS searches at 13 TeV and 36.1 fb$^{-1}$ of data. Three search channels are investigated: di-lepton production, four lepton production and di-lepton production plus missing transverse energy. These channels are used to derive exclusion limits on $Z^\prime_{331}$, doubly-charged and singly-charged bileptons, respectively. A $Z^\prime_{331}$ with mass smaller than 3.7 TeV is excluded. For bileptons, the coupling strength to leptons is varied and its found that doubly-charged bileptons with masses up to 1.2 TeV are excluded. For singly-charged bileptons, the maximum bound is 850 GeV, and it represents the very first direct limit on singly-charged bileptons obtained with hadron collision data. These results improves previous limits on 331 vector bosons and at the moment this paper is written, they are the most stringent limits on these particles. 

\section*{acknowledgements}
This work is partially supported by the Brazilian National Council for Scientific and Technological Development (CNPq) under grants 308494/2015-6 and 402846/2016-8.
%\bibliographystyle{apsrev4-2}
%\bibliography{bibliography}

\end{document}